\documentclass{ws-p8-50x6-00}
\usepackage[dvips]{color}                 %%% for colors in the graphs
\usepackage{amssymb}
\usepackage{amsmath}

\usepackage{xspace}                       %%% to get the correct spacing of
\newcommand{\ia}{{\"{\i}}}   %not necessary if \usepackage[T1]{fontenc} is used
\newcommand{\PR}{\textit{Phys.\ Rev.\ }}

\newcommand{\PRC}{\PR\textbf{C}}
\newcommand{\kv}{\vec{k}}

\newcommand{\MeV}{\mathrm{MeV}}

\newcommand{\fm}{\mathrm{fm}}

\newcommand{\HBChiPT}{HB$\chi$PT\xspace}

\begin{document}

\title{Learning from Dispersive Effects in the Nucleon Polarisabilities}

\author{Harald W.\ Grie\3hammer}
  
\address{T39, Technische Universit{\"a}t M{\"u}nchen, D-85747 Garching,
  Germany \\EMail:
  hgrie@physik.tu-muenchen.de\footnote{Talk held at Baryons 2002, JLab,
    Newport News (USA), 4-8 March 2002. To be published in the proceedings.
    Preprint numbers nucl-th/0204073, TUM/T39-02-06.}
  }


\maketitle

%\absatracts{A cartoon is drawn how and why to define and extract the energy
%  dependence of the nucleon polarisabilities from low energy Compton
%  scattering on the nucleon. The information dynamical polarisabilities
%  contain about the low energy degrees of freedom inside the nucleon is
%  extracted by comparing the chiral effective field theory calculation with
%  dispersion theory results.  Dynamical polarisabilities do not contain more
%  or less information than the corresponding Compton scattering amplitudes,
%  but as with any multipole decomposition, the facts are more readily
%  accessible and easier to interpret. Stringent constraints for models and
%  model-independent power countings of a low energy effective field theory of
%  the nucleon follow from their analysis.}

\noindent
Static nucleon polarisabilities gauge the stiffness of the nucleon against an
external electro-magnetic field, parameterising the part of the real Compton
amplitude at zero energy which is not explained by the pole terms, i.e.~by the
successive interactions of two photons with a point-like nucleon of anomalous
magnetic moment $\kappa$. \emph{Dynamical} nucleon polarisabilities are the
energy dependent generalisation to real Compton scattering and thus provide
more information about the low energy effective degrees of freedom inside the
nucleon. Here, I sketch their definition and interpretation in terms of the
low energy degrees of freedom, referring to~\cite{hgth,pols2} for particulars
and references.

Investigating in more detail the nucleon polarisabilities at non-zero photon
energy, one first subtracts the ``nucleon pole'' effects from the real Compton
amplitude $T$, writing $ \bar{T}(\omega,z)=\bar{A}_1(\omega,z)\,(
\vec{\epsilon}^{\,\prime\ast}\cdot\vec{\epsilon})+ \bar{A}_2(\omega,z)\;
\left(\vec{\epsilon}^{\prime\,\ast}\cdot\hat{\kv}\right)
\left(\vec{\epsilon}\cdot\hat{\kv}{}^\prime\right)$\newline$+\ldots\;, $ for the two
spin-independent structure amplitudes in the centre of mass (cm) frame, with
the non-Born part already subtracted. $\omega$ is the cm energy of an incident
(outgoing) real photon with momentum $\kv$ ($\kv^\prime$) and polarisation
$\vec{\epsilon}$ ($\vec{\epsilon}^{\prime}$), scattering under the cm angle
$\theta$ off the nucleon, with $z=\cos\theta$.

%Clearly, the differential cross sections are independent of
%this artificial separation of the amplitudes into ``pole'' and ``non-pole''
%parts.

At fixed real photon energy, the structure dependent part of the amplitude is
then analysed by expanding it in multipoles. In terms of the first four
electric and magnetic (dipole and quadrupole) polarisabilities of definite
multipolarity,
$\alpha_{E1}(\omega),\,\beta_{M1}(\omega),\,\alpha_{E2}(\omega)$ and
$\beta_{M2}(\omega)$, the amplitudes read
\begin{eqnarray}
  \label{eq:amplitudes1}
  \bar{A}_1(\omega,z)&=&\frac{4\pi\,W}{M}\Bigg[
  \Big(\alpha_{E1}(\omega)+z\beta_{M1}(\omega)\Big)\omega^2+\nonumber\\
  &&\;\;\;\;\;\;\;\;\;\;\;+\;
  \frac{1}{12}\Big(z\alpha_{E2}(\omega)+(2 z^2-1)\beta_{M2}(\omega)
  \Big)\omega^4+\dots\Bigg]
  \\
  \bar{A}_2(\omega,z)&=&-\;\frac{4\pi\,W}{M}
  \Bigg[
  \beta_{M1}(\omega)\;\omega^2+
  \frac{1}{12}\Big(-\alpha_{E2}(\omega)+2 z\beta_{M2}(\omega)
  \Big)\omega^4+\dots\Bigg]\nonumber
  \;\;,
\end{eqnarray}
where $W=\omega+\sqrt{M^2+\omega^2}$ is the total cm energy and $M$ the
nucleon mass.  The multipolarities are dis-entangled by their angular
dependence.  The pre-factors are chosen such that at zero photon energy, the
definitions of the static polarisabilities are recovered, e.g.: $
\bar{\alpha}_E=\alpha_{E1}(\omega=0),\; \bar{\beta}_M=\beta_{M1}(\omega=0)$.

\begin{figure}[!htb]
  \begin{center}
    \includegraphics*[width=0.87\textwidth]{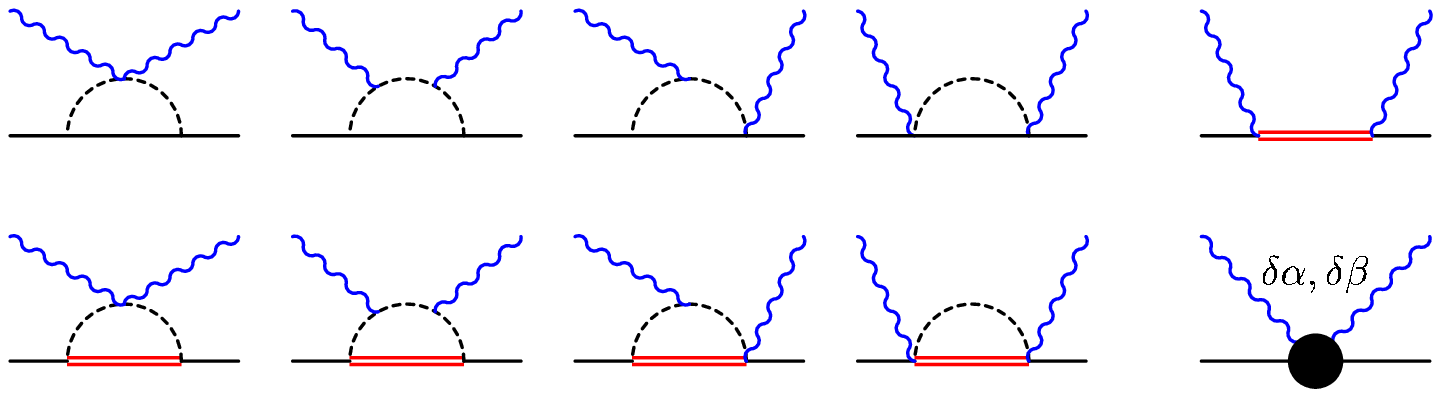}
    \caption{The diagrams contributing at leading one loop order 
      in MSSE. Graphs obtained by permuting vertices or external lines are not
      displayed. Double line: $\Delta(1232)$.}
    \label{fig:MSSEdiagrams}
  \end{center}
\end{figure}

\begin{figure}[!htb]
  \begin{center}
    \includegraphics*[width=0.45\textwidth]{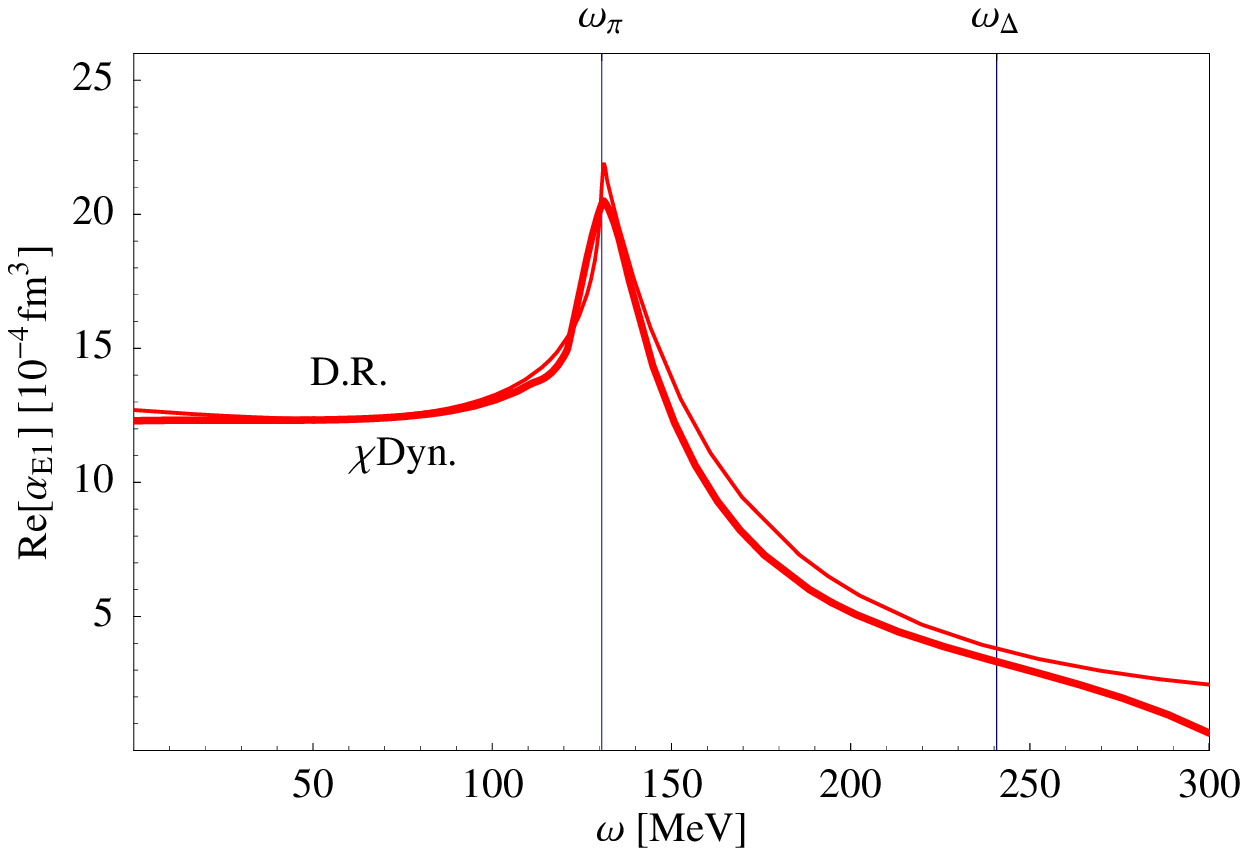}
    \includegraphics*[width=0.45\textwidth]{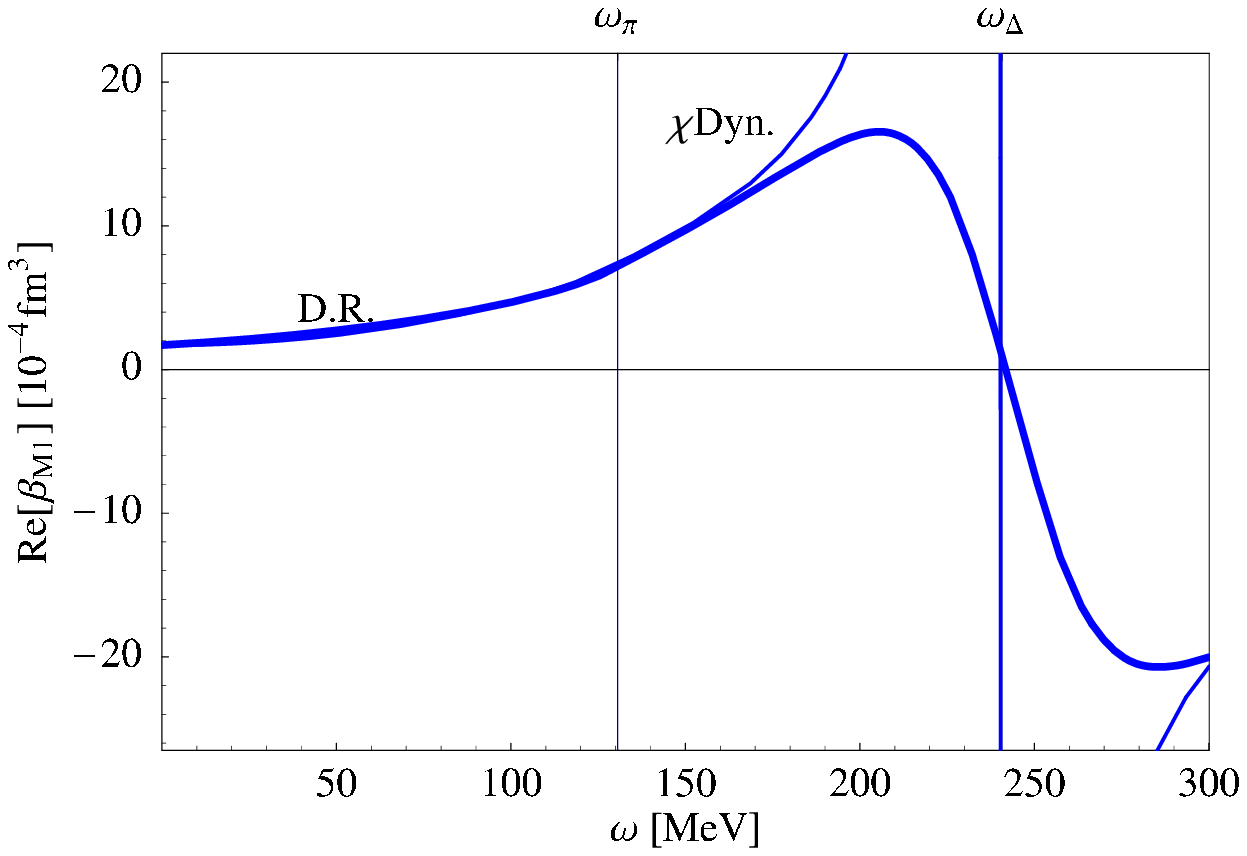}
    \\
    \includegraphics*[width=0.45\textwidth]{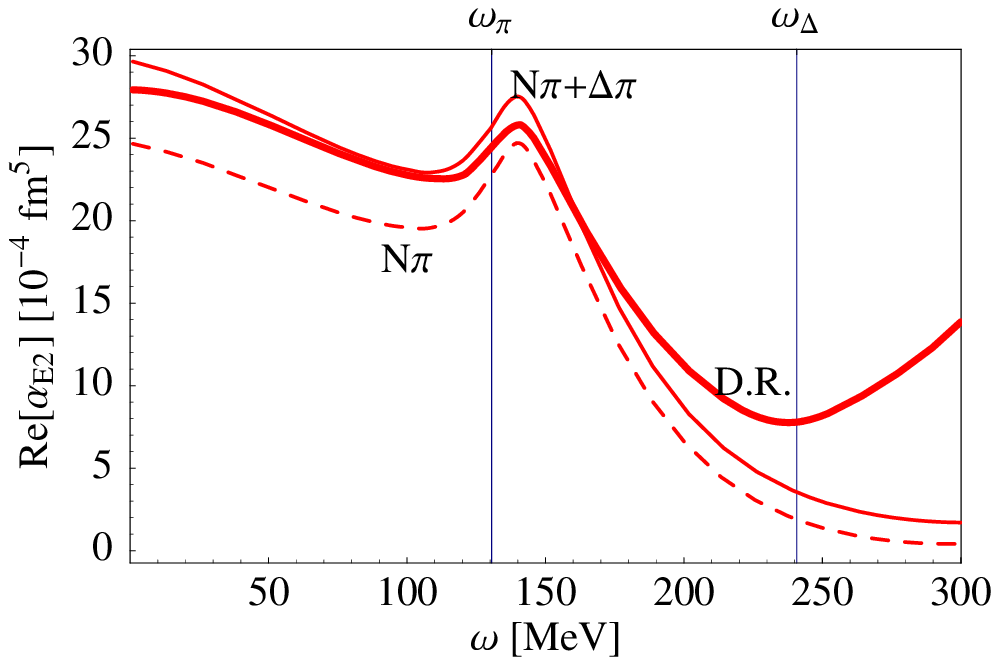}
    \includegraphics*[width=0.45\textwidth]{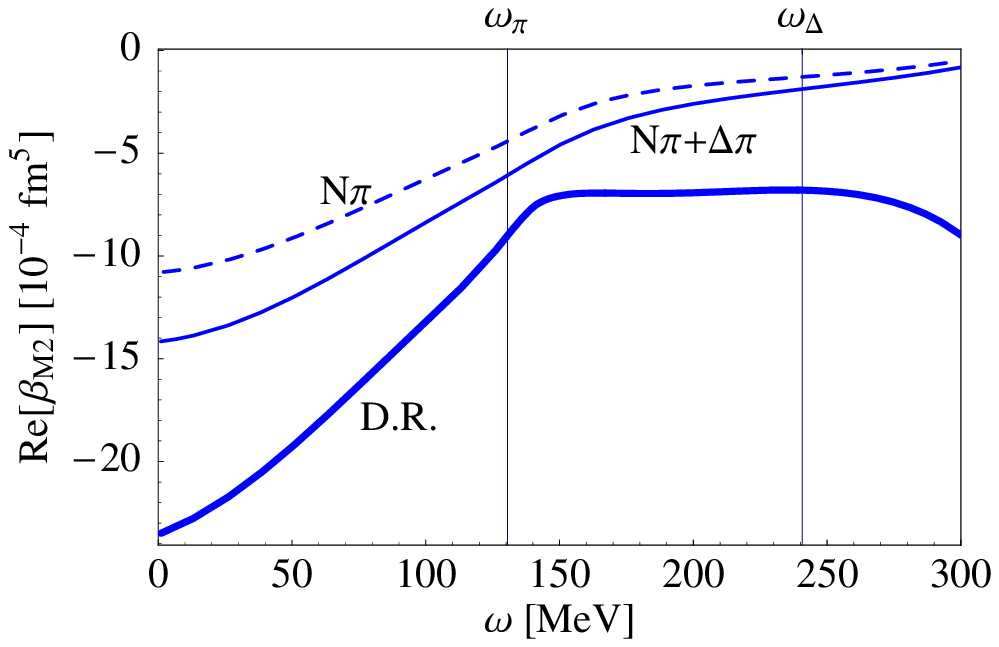}
    \\
    \includegraphics*[width=0.44\textwidth]{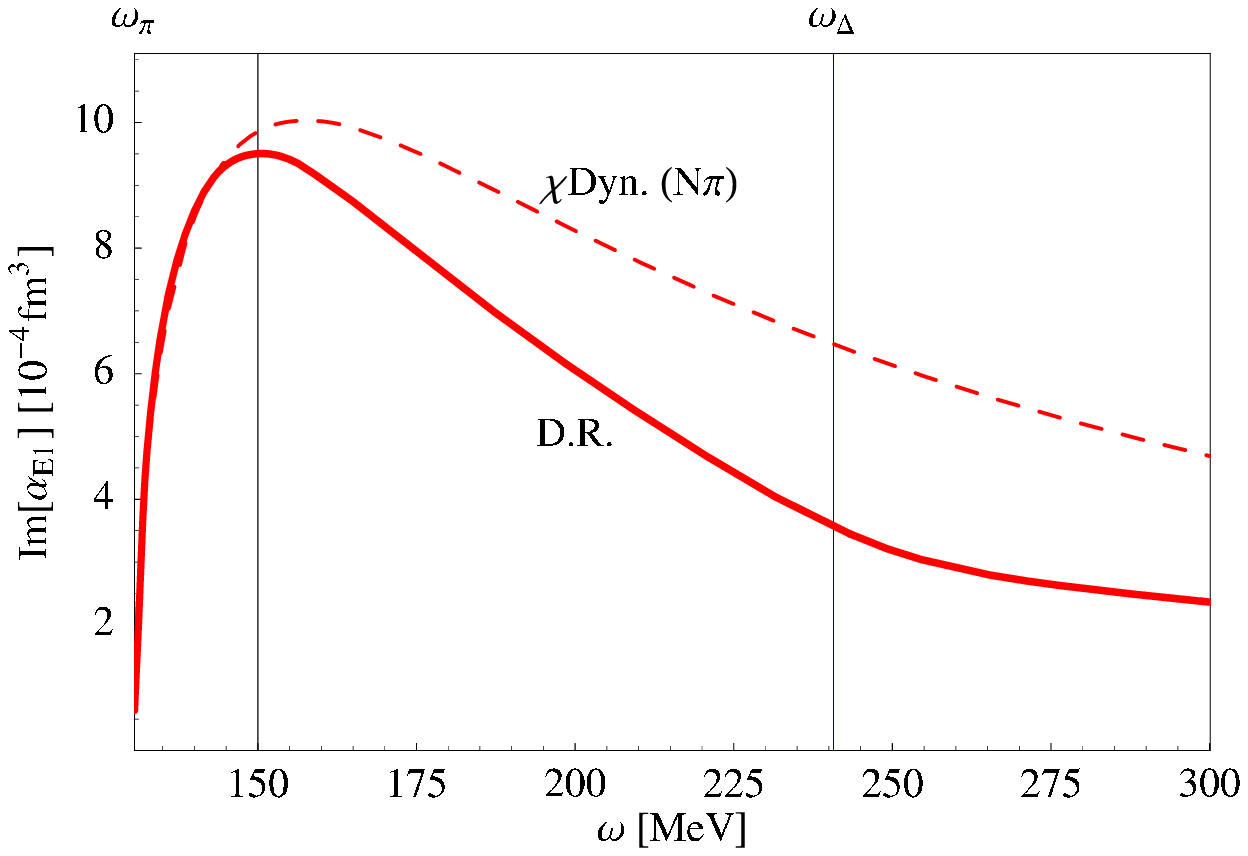}
    \includegraphics*[width=0.44\textwidth]{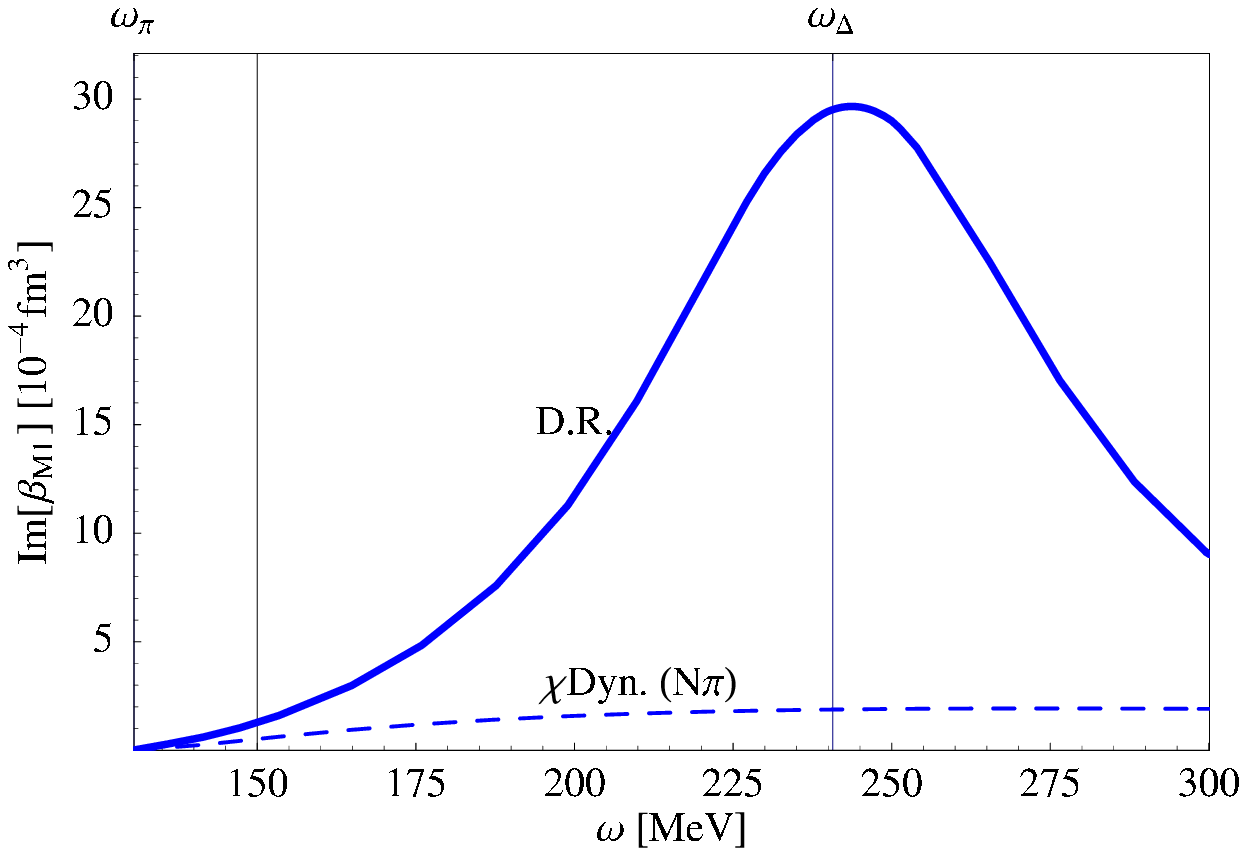}
    \\
    \includegraphics*[width=0.44\textwidth]{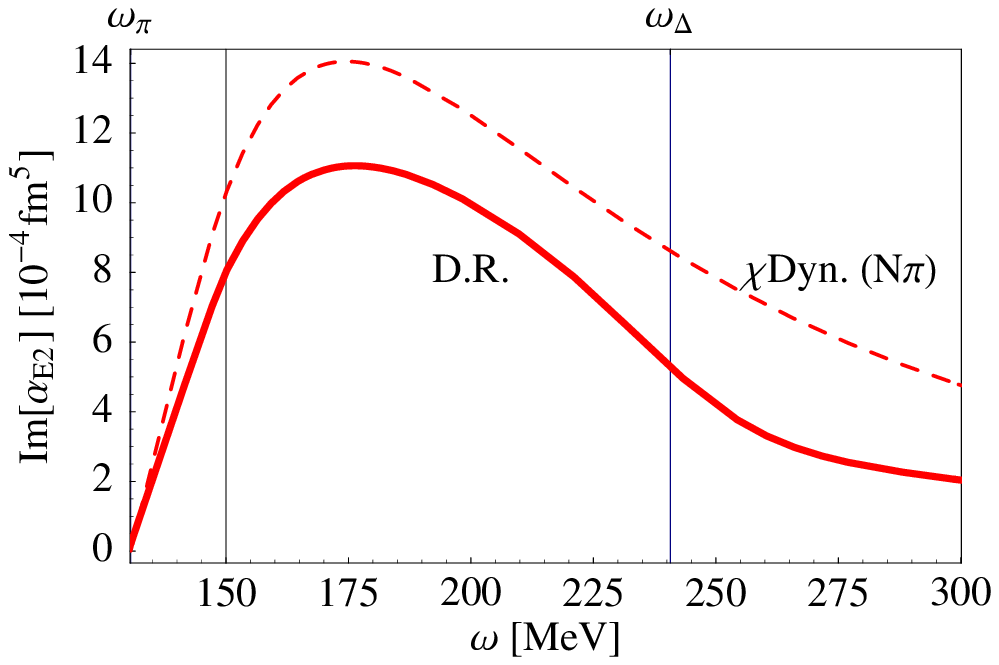}
    \includegraphics*[width=0.44\textwidth]{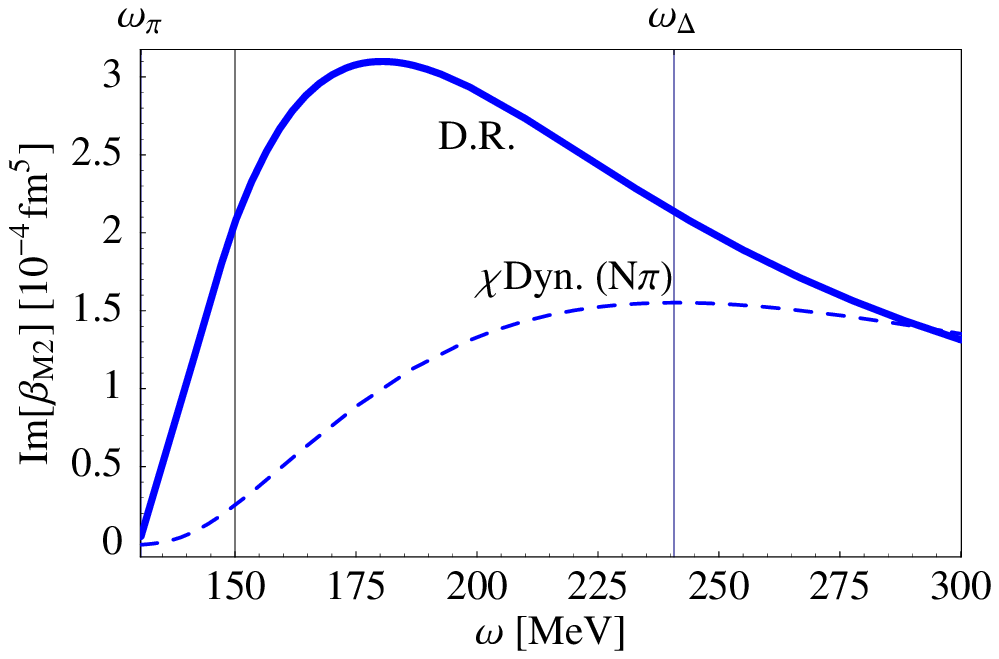}
    \caption{Comparing the Dispersion Relations results (thick solid line)
      of the real (top four) and imaginary (bottom four) parts of the
      iso-scalar dynamical electric and magnetic dipole and quadrupole
      polarisabilities with the leading one loop order MSSE prediction.
      Dashed: HB$\chi$PT prediction (nucleons and pions only); dotted:
      $\Delta$ pole contribution added; dot-dashed: pion cloud around $\Delta$
      added; thin solid: total MSSE result (counter terms for the dipole
      polarisabilities fixed to the static values of the dipole
      polarisabilities). For the imaginary parts, the MSSE and \HBChiPT
      predictions are identical at this order.}
    \label{fig:totalpolsresult}
  \end{center}
\end{figure}

Dynamical polarisabilities test the temporal response of the global, low
energy excitation spectrum of the nucleon at non-zero energy. They therefore
contain information about dispersive effects induced by internal relaxation
mechanisms, baryonic resonances and meson production thresholds of the
nucleon. This is also clearly seen in Figs.~\ref{fig:totalpolsresult}, where
the result of a dispersion theory analysis of the four leading dynamical
polarisabilities is compared to a chiral effective field theory describing the
nucleon at low energies, called Modified Small Scale Expansion MSSE.  The
latter is a rigorous, model-independent and systematic approach to low energy
QCD. It incorporates the nucleon, the $\Delta(1232)$ and their respective pion
clouds as explicit low energy degrees of freedom, see
Fig.~\ref{fig:MSSEdiagrams} for the dominant contributions~\cite{pols2}.
Short distance physics is sub-sumed into local counter terms whose strengths
are na{\ia}vely given by dimensional analysis to be of the order of
$|\delta\alpha_{E1},\delta\beta_{M1}|\sim1-2$\footnote{All dipole
  polarisabilities here given in the ``natural units'' of $10^{-4}\;\fm^3$.}
which would make them higher order effects. However, fitting these two free
parameters to reproduce the static values of the dipole polarisabilities shows
that they are indeed much larger, $\delta\alpha_{E1}\approx
-5,\;\delta\beta_{M1}\approx-10$, i.e.~of leading order in accord with the
MSSE power counting. It is at this point that the new dimension of the
dynamical polarisabilities comes into play by allowing for a
\emph{parameter-free prediction} of the energy dependence after the zero
energy value is fixed.

For example, a large dia-magnetic but only very weakly energy dependent
contribution is needed in the magnetic dipole polarisability to cancel the
well known large para-magnetism coming from the strong $M1\to M1$ transition
between nucleon and $\Delta$ which clearly rules this channel.  As the shape
of $\beta_{M1}(\omega)$ is dominated by the $\Delta$ already below the pion
production threshold $\omega_\pi$, the good agreement between its
experimentally measured value at zero energy ($\beta_{M1}(0)=1.5$) and the
result of Heavy Baryon Chiral Perturbation Theory \HBChiPT with only pions and
nucleons as low energy degrees of freedom
($\beta_{M1}^{\mathrm{HB}\chi\mathrm{PT}}(0)=1.2$) can be seen as accidental.
The energy dependence of the dynamical polarisabilities demonstrates therefore
that the underlying physics mimicked at zero energy by the counter term is in
a large range insensitive to derailed dynamics at short distances. No
genuinely new degrees of freedom are missed at low energies. As the $\Delta$
has no width at leading order in MSSE, the result for $\beta_{M1}(\omega)$
diverges close to the $\Delta$ resonance energy $\omega_\Delta$. For this
reason, the imaginary part of $\beta_{M1}(\omega)$ above the one pion
production threshold is also ill reproduced in MSSE.  Dispersion theory shows
that it is clearly dominated by the non-zero width of the $\Delta(1232)$.  The
cusp strength and structure at $\omega_\pi$ is well captured in MSSE, and very
well so for $\alpha_{E1}(\omega)$.

Thanks to the contribution from the pion cloud around the $\Delta$, the
agreement for $\alpha_{E2}(\omega)$ at low energies is good. NOt surprisingly,
there is no residue of the famed $E2 \rightarrow E2$
transition due to its smallness. No counter term is used or necessary in this
channel to account for short distance physics. A 
nucleon resonance might be seen in the discrepancy between the DR and chiral
calculations at very high momenta $\omega>250\;\MeV$.

Albeit there is some improvement by adding the $\Delta\pi$ continuum, nearly
half of the static strength of $\beta_{M2}(\omega)$ is still missing.
Introducing a counter term to reproduce the correct static value, as was done
for the dipole polarisabilities but is for the quadrupole polarisabilities
inconsistent with the MSSE power counting, is of no help in the magnetic
quadrupole polarisability because the slope of the dispersion theory result is
much steeper than the one obtained in MSSE. Not even the physics of the pion
production threshold seems to be reproduced correctly.

This analysis is an overall success for MSSE: In the r\'egime at low energies
where it is supposed to work, the agreement is surprisingly good, and the
deviation around $\omega_\Delta$ is clearly related to the fact that the
``natural'' power counting must be modified in order to account for the
non-zero width of a resonance by summing a sub-class of diagrams.

Dynamical polarisabilities do not contain more or less information than the
corresponding Compton scattering amplitudes, but as with any multipole
decomposition, the facts are more readily accessible and easier to interpret.
Stringent constraints for models and model-independent power countings of a
low energy effective field theory of the nucleon follow from their analysis.

\begin{thebibliography}{99}
\bibitem{hgth} H.~W.~Grie\3hammer and T.~R.~Hemmert: \PRC\textbf{65}, 045207.
\bibitem{pols2} R.~Hildebrandt, H.~W.~Grie\3hammer, T.~R.~Hemmert and
  B.~Pasquini, forthcoming.
  
\end{thebibliography}
\end{document}